\documentclass[conference]{IEEEtran}
\IEEEoverridecommandlockouts

\usepackage{cite}
\usepackage{amsmath,amssymb,amsfonts}
\usepackage{algorithmic}
\usepackage{graphicx}
\usepackage{textcomp}
\usepackage{tabularx}
\usepackage[table,xcdraw]{xcolor}
\usepackage{url}

\def\BibTeX{{\rm B\kern-.05em{\sc i\kern-.025em b}\kern-.08em
    T\kern-.1667em\lower.7ex\hbox{E}\kern-.125emX}}
\begin{document}

\title{OpenEnded: An Open-Response Speech Corpus for Speaking Proficiency Assessment with Human Annotations and ALM Supervision
}

\author{
\IEEEauthorblockN{Yu-Wen Chen\qquad Eric Zhou\qquad Evelyn Ding\qquad Tianyi Shen\qquad Zhou Yu\qquad Julia Hirschberg}
\IEEEauthorblockA{
Department of Computer Science, Columbia University, USA
}
}

\maketitle

\begin{abstract}
The development of automated speaking assessment (ASA) is limited by the scarcity of public datasets, with most existing work relying on read-aloud speech, which limits applicability to real-world communication scenarios. In this work, we introduce OpenEnded, a corpus of English practice speech from Mandarin speakers in open-response tasks. Unlike prior open-response datasets that provide only holistic proficiency scores, OpenEnded offers utterance-level assessments of accuracy, fluency, and prosody. Approximately 10,000 utterances are collected and annotated using a hybrid framework: 1,000 are manually labeled via multi-rater scoring with discrepancy resolution to form a high-quality test set, while the remaining are pseudo-labeled by an audio language model (ALM) for training and development sets. We evaluate ALMs and existing ASA models on the OpenEnded test set and introduce VoxPA as an additional baseline. Results show that ALM-generated pseudo-labels improve training over original ALM scoring, while VoxPA achieves the best performance among all baselines.

\end{abstract}


\section{Introduction}
Automated speaking assessment (ASA), which provides automatic feedback to language learners, can improve the accessibility and affordability of language learning \cite{eskenazi2009overview, kheir2023automatic, rogerson2021computer}. To enable the development of ASA systems, diverse annotated speech datasets reflecting real-world learning are essential. However, existing open-source datasets remain limited, and many prior studies have relied on read-aloud datasets in which speakers read from a script. One of the most widely used datasets is Speechocean762~\cite{zhang2021speechocean762} (hereafter referred to as Speechocean), a read-aloud dataset that provides predefined target sentences together with utterance-level assessments of accuracy, fluency, and prosody. Based on this dataset, researchers have developed models that rely on target sentences as input~\cite{yan2025conpco}, including Goodness-of-Pronunciation (GOP)-based models~\cite{liu2023phone,gong2022transformer, sheoran2023pronunciation, do2023hierarchical, yan2024effective}. However, since these models are trained using target sentences, their performance may degrade substantially in open-response scenarios, where learners produce free-form responses that more closely reflect real-world communication. Other studies have explored approaches that either use automatic speech recognition systems to generate perceived transcripts as replacements for the target transcripts~\cite{chen2023multipa}, or avoid using transcripts entirely~\cite{liu2023asr}. Although these models should be applicable to open-response settings, their training and evaluation are still primarily conducted on read-aloud datasets, demonstrating the impact of limited data availability on ASA development.

A limited number of datasets containing open-response speech have been released to the public. For example, Speak \& Improve~\cite{knill2024speak} contains multi-level monologue tasks annotated with CEFR (Common European Framework of Reference for Languages) levels, which provide a holistic measure of language proficiency. The ICNALE~\cite{ishikawa2019icnale} dataset consists of oral interview recordings and categorizes speakers into CEFR-based proficiency levels. However, to the best of our knowledge, no existing publicly available dataset for open-response settings provides utterance-level feedback on accuracy, fluency, and prosody. These three metrics, which together provide a comprehensive evaluation of a speaker’s ability, enable more fine-grained and actionable assessment by distinguishing specific strengths and weaknesses for each utterance. 

The lack of such datasets may stem from the challenge that human annotation is both time-consuming and expensive. This motivates us to use pseudo-labeling as an alternative to reduce manual annotation effort, where model-generated labels are used to supervise downstream learning. Such an approach has been shown to be effective in prior work, where pseudo-labels generated by large language models (LLMs) can improve downstream task performance~\cite{wang2021want, gilardi2023chatgpt, ding2023gpt, he2024annollm}. More recently, audio language models (ALMs), which extend LLMs to speech inputs, have demonstrated emerging capabilities in speech understanding~\cite{chu2024qwen2, chen2025read, arora2025landscape}. However, their use for pseudo-labeling in ASA remains underexplored. 

In this study, we introduce OpenEnded\footnote{Dataset and model: \url{https://github.com/yuwchen/OpenEnded}.}, a corpus of speakers practicing English in open-response scenarios, annotated with utterance-level measures of accuracy, fluency, and prosody. To build this dataset, we first collected audio recordings of Mandarin speakers practicing English with a chatbot. We then sampled a subset of the collected data, designed a rubric with annotation criteria for each assessment dimension (i.e., accuracy, fluency, and prosody), and manually annotated it to form the OpenEnded test set. The manual annotation process consisted of independent scoring from three human annotators, followed by discrepancy resolution through discussion, continuous rubric refinement, and expert consultation to ensure annotation quality and consistency. Among the approximately 10,000 collected recordings, 1,000 samples were manually annotated to form the OpenEnded test set, while the remaining samples were pseudo-labeled by ALMs to construct the training and development sets.

We report benchmark results on the OpenEnded test set using representative ALM and ASA models, and introduce VoxPA as part of the benchmark. Experimental results demonstrate that ALM-generated pseudo-labels (i.e., the OpenEnded training and development set) can effectively support ASA model training, achieving better performance than the original ALM-based scoring. Moreover, VoxPA further improves over previous ASA models, demonstrating the effectiveness of the proposed architecture and providing a stronger baseline for the OpenEnded dataset.

\section{Methodology}

\begin{figure}[tb]
\centerline{\includegraphics[scale=0.92]{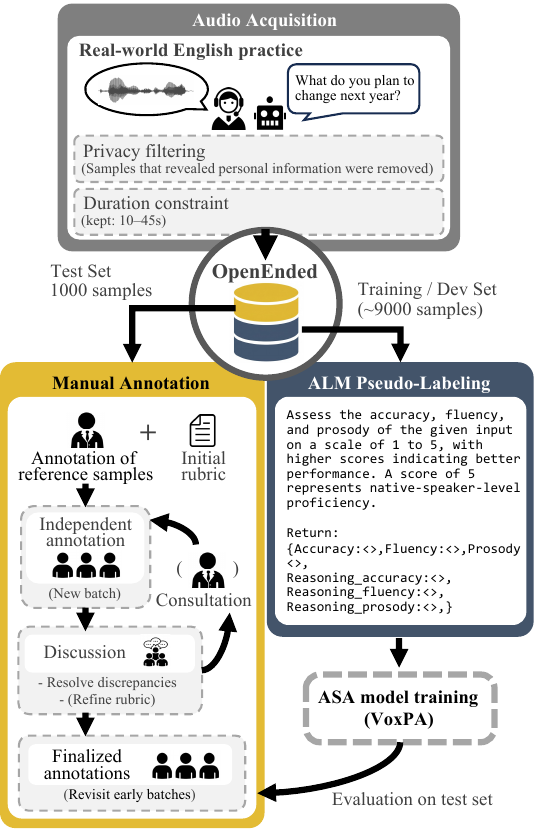}}
\caption{Study overview.}
\label{fig:openended_overview}
\end{figure}

Table~\ref{tab:dataset} compares OpenEnded with existing public speaking proficiency assessment datasets. OpenEnded differs from prior work by its open-response setting and fine-grained utterance-level annotations of accuracy, fluency, and prosody on a 1–5 (poor to excellent) scale. It includes a manually annotated test set and a pseudo-labeled training and development set. To evaluate the effectiveness of pseudo-labeling, we train ASA models on the pseudo-labeled data and measure performance gains on the test set. We further introduce VoxPA, a new baseline ASA model. Fig.~\ref{fig:openended_overview} provides an overview of the study. The following sections detail the dataset construction pipeline and the VoxPA model.

\begin{table*}[t]
\caption{Comparison of OpenEnded with Public Speaking Proficiency Assessment Datasets}
\begin{center}
\renewcommand{\arraystretch}{1.1}
\scalebox{0.88}{
\begin{tabularx}{1.1\textwidth}{l|X|X|X|X}
\hline \hline
Dataset   & Scenario                               & Native Language / \newline Country of Origin          & \# of Utterances                  & Scoring Scheme            \\ \hline \hline
L2-ARCTIC~\cite{zhao2018l2} & Read-aloud                             & Hindi, Korean,  \newline Mandarin, Spanish, and Arabic & 11,026 (1,499 manually annotated) & Mispronunciation errors   \\ \hline
Speechocean~\cite{zhang2021speechocean762} &
  Read-aloud &
  Mandarin &
  5,000 &
  Utter-Acc., Flu., Pros. \newline Word-Acc., Stress \newline Phoneme-Acc. \\ \hline
Speak \& Improve~\cite{knill2024speak} & 
 Read-aloud +\newline Elicited speech &
  Global & $\sim$50,000 (training set partially annotated) &
  CEFR holistic scale \\ \hline
ICNALE~\cite{ishikawa2019icnale}    & Elicited speech & Asian, Native English                        & 4,400     &  CEFR-based holistic scale \\ \hline \hline
 OpenEnded & Elicited speech &
  Mandarin &
  9,782 (1,000 manually annotated with the remainder pseudo-labeled) &
  Utter-Acc., Flu., Pros. \\ \hline \hline
\end{tabularx}
}
\label{tab:dataset}
\end{center}
\end{table*}

\subsection{OpenEnded Development Process}

\subsubsection{Audio Acquisition}

We collected audio from native Mandarin speakers practicing English using a dialog-based chatbot in an open-response setting. Specifically, speakers were presented with questions such as “What do you plan to change next year?” and were instructed to respond freely. As a real use case for ASA, speakers access the system with their own headsets at their own location. The collected data include the audio recording of their response, the question asked, and the speaker’s ID.

The collected audio recordings were passed through a privacy filtering process, as speakers may inadvertently disclose personal information in their responses. The filtering pipeline includes: (1) a named entity recognition (NER)–based module that identifies mentions of persons, organizations, geopolitical entities, nationalities, and facilities; and (2) an LLM model that evaluates each audio’s transcript using the prompt: “Check if the transcript contains private information.” Only recordings that pass both filtering stages were retained. After privacy filtering, we further applied a duration constraint, retaining only recordings between 10 and 45 seconds. Then we randomly sampled 1,000 utterances from 250 distinct speaker IDs to construct the test set. For the remaining recordings, we excluded any utterances whose speaker IDs overlap with those in the test set. These samples were then split into training and development sets containing 6,109 and 2,673 audio recordings, respectively. Together, the training and development sets included recordings from 753 distinct speaker IDs.

\subsubsection{Manual Annotation}
\begin{table*}[htbp!]
\caption{Final Annotation Rubric}
\begin{center}
\renewcommand{\arraystretch}{1.1}
\scalebox{0.88}{
\begin{tabularx}{1.1\textwidth}{l|X|X|X}
\hline \hline
 & Accuracy  & Fluency  & Prosody               \\ \hline \hline
1 - Very Poor & \textbf{Grammatical errors} are extensive, affecting the fundamental speaking flow and patterns. \textbf{Pronunciation} of individual words is also very poor. \textbf{Vocabulary} is clearly limited, making it impossible to understand what is being said.                             & \textbf{Flow} of speech is halting, fragmented, and slow. Very long \textbf{pauses}, excessive \textbf{filler words} (e.g., um, uh), or lots of repetition and restarting. The speaker often seems to search for words, essentially stuck and unable to finish an idea. & \textbf{Intonation} is monotone or highly unnatural. The majority of words are incorrectly pitched. \textbf{Stress} is almost always misplaced or absent. \textbf{Rhythm} is flat or erratic. The speech is very uncomfortable to listen to.  \\ \hline
2 - Weak &
  \textbf{Grammatical errors} occur frequently throughout most sentences. \textbf{Pronunciation} is poor, making it difficult to understand what is being said. \textbf{Vocabulary} is limited and affects the flow of speech. &
  \textbf{Flow} is interrupted by noticeable hesitation and uneven pacing. Frequent \textbf{pauses} or \textbf{filler words} between ideas. Delivery feels effortful, but not quite stuck. &
  \textbf{Intonation} is off and not accurate. The speaker \textbf{stresses} words in a way that is either monotone or exaggerated in the wrong direction, making it feel unnatural. \textbf{Rhythm} is odd and uncomfortable. \\ \hline
3 - Fair&
  \textbf{Grammatical errors} are quite common across multiple aspects, often with mistakes in verb tense or word usage. \textbf{Pronunciation} is fine, with only a few words being spoken incorrectly. \textbf{Vocabulary} is not particularly impressive. &
   \textbf{Flow} has moderate hesitations. The speaker sometimes \textbf{pauses} to think about their next word, leading to abrupt silence, or they use \textbf{filler words} to fill the space between words. The speaker may either speak slowly and make hesitations, or speak rapidly and struggle to sustain their pace.
 & \textbf{Intonation} has solid variation. However, syllables may be incorrectly \textbf{stressed} or oddly paced. \textbf{Rhythm} is off. The speaker feels as if words are blurted out as soon as they think of them, instead of placing them in the context of a naturally inflected sentence.   \\ \hline
4 - Good & \textbf{Grammatical errors} are mainly prevalent in one aspect (e.g. verb tense), or there are only a few grammatical errors across different aspects. \textbf{Pronunciation} is good overall. \textbf{Vocabulary} is relatively diverse.   & \textbf{Flow} is generally smooth. Any \textbf{pauses} are brief and the speaker recovers quickly, and the use of \textbf{filler words} is limited. Overall, the speaker's pacing is natural and easy to listen to.                        & \textbf{Intonation} cues are mostly correct and feel expressive in a natural way. \textbf{Stress} on syllables may occasionally be incorrect or monotone. \textbf{Rhythm} is reasonable, and emphasis aligns well with meaning.                             \\ \hline
5 - Excellent & \textbf{Grammatical errors} are not present, and the speaker achieves consistently accurate grammar and syntax. \textbf{Pronunciation} feels native-level. \textbf{Vocabulary} reflects a deeper understanding of the English language, and the speaker's message is clear and precise. &
  \textbf{Flow} feels smooth and effortless. There are very few, if any, \textbf{pauses} and \textbf{filler words}. The speaker's control over their pacing is reflected in confident delivery. &
  \textbf{Intonation} is correctly expressed with a large range in pitch. \textbf{Stress} is accurately placed. \textbf{Rhythm} is good. The speaker's fully natural prosody enhances expressiveness and comprehension.  \\ \hline
\end{tabularx}
}
\label{tab:annotation_rubric}
\end{center}
\end{table*}

Three annotators with high English proficiency performed the annotations, with a senior expert in English and speech assessment providing continuous calibration support throughout the annotation process. First, the senior expert established reference scores on a subset of samples, which served as calibration guidelines for the annotators. The annotators established an initial annotation rubric and independently scored each utterance. Samples in which any of the three ratings (accuracy, fluency, or prosody) differed by two or more points were jointly reviewed and revised through discussion to reach consensus. During this process, the rubric was iteratively refined. Throughout the annotation process, the expert was consulted whenever annotators were uncertain about how to score specific samples. After completing the full annotation process, the first one-third of the samples were revisited and updated to improve consistency. The final annotation rubric is presented in Table~\ref{tab:annotation_rubric}.

\subsubsection{ALM pseudo-labeling}

The ALM was chosen for pseudo-labeling because (1) it does not require additional speaking proficiency assessment data for training, (2) it benefits from large-scale text corpora, which provide linguistic knowledge relevant to different aspects of speaking proficiency assessment~\cite{chen2025read, wang2023assessing}, and (3) it generates descriptive feedback along with scores, providing interpretable guidance for learner improvement. The ALM was prompted to perform multi-task ASA by jointly predicting accuracy, fluency, and prosody scores on a 1–5 scale, along with providing reasoning for the assigned scores (Fig.~\ref{fig:openended_overview}). Note that the manual annotation rubric is excluded from the prompt, as our pilot study shows that it does not improve performance while substantially increasing prompt length.

\subsection{VoxPA}

\begin{figure}[htbp!]
\centerline{\includegraphics[scale=0.83]{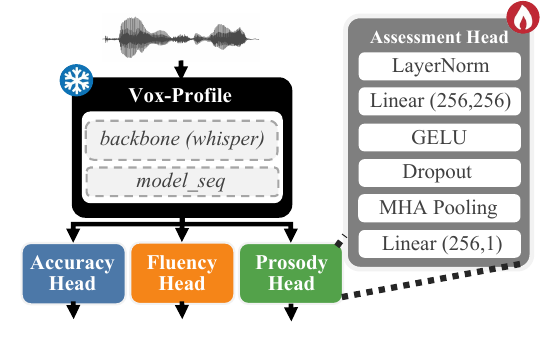}}
\caption{VoxPA architecture. \emph{Backbone} and \emph{model\_seq} are original modules from Vox-Profile.}
\label{fig:vox_pa}
\end{figure}

We also release VoxPA (Fig.~\ref{fig:vox_pa}) as a dataset-specific baseline, built upon Vox-Profile~\cite{feng2025vox}, a Whisper-based speech model for characterizing diverse speaker and speech attributes. VoxPA extends Vox-Profile with separate assessment heads for accuracy, fluency, and prosody. Each head comprises a LayerNorm layer, a linear projection with GELU and dropout, a multi-head attention (MHA) pooling module, and a final linear layer for utterance-level score prediction. During training, the Vox-Profile modules are frozen and only the assessment heads are optimized. 

\section{Experimental Setup}

\subsection{Model and Dataset Construction}

For the privacy filtering pipeline, NER was performed using the spaCy \emph{en\_core\_web\_sm} model to identify entities belonging to the categories \emph{PERSON}, \emph{ORG}, \emph{GPE}, \emph{NORP}, and \emph{FAC}. The LLM-based filtering stage and ALM pseudo-labeling were performed using \emph{Gemini-2.0-Flash} (hereafter Gemini-2.0) with default settings. Gemini~\cite{comanici2025gemini} was selected as a recent leading large-scale multimodal foundation model with capabilities for handling audio inputs and multimodal understanding. For VoxPA, the Vox-Profile was initialized from the \emph{tiantiaf/whisper-large-v3-voice-quality} checkpoint from \emph{huggingface}. The model was trained using a weighted mean-squared error loss, where higher weights were assigned to samples whose label scores deviate further from the median of the training set. Optimization was performed using SGD with a learning rate of 5e-5 and a momentum of 0.7. Early stopping with a patience of 2 epochs was applied, and the checkpoint achieving the best validation performance was retained. During inference we applied a rolling-window strategy for audio longer than 15 seconds, as the maximum training duration of Vox-Profile is 15s. Specifically, we used a 15-second window with a 7.5-second hop size, and averaged the predictions across all windows to obtain the final output.

\subsection{Evaluation Setup}

We report benchmark results on the OpenEnded test set using representative ALM and ASA models. For ALM-based methods, we evaluate Gemini models, including a multi-task configuration (\emph{gemini-2.0-AFP} and \emph{gemini-2.5-AFP}), where the model is run once to jointly predict accuracy, fluency, and prosody (Fig.~\ref{fig:openended_overview}), as well as single-task variants (\emph{gemini-2.0-A}, \emph{-F}, and \emph{-P}), which independently predict accuracy, fluency, and prosody, respectively. The single-task prompts follow the same formulation as the multi-task setting, with the only modification being the specification of one target dimension. For analysis of the reasoning in Sec.~\ref{sec:exp_alm_asa}, sentence embeddings are computed using \emph{BAAI/bge-large-en-v1.5}~\cite{bge_embedding}.

GOPT~\cite{gong2022transformer}, MultiPA~\cite{chen2023multipa}, and Joint-APA~\cite{ryu2023joint} were selected as baselines because their architectures take speech recordings as input and predict utterance-level assessments of accuracy, fluency, and prosody, matching the rating dimensions of OpenEnded and targeting speaking proficiency assessment. For GOPT, we used the Librispeech checkpoint released in the original GitHub repository. Since GOPT relies on GOP features that require ground-truth transcripts, which are not available in the OpenEnded dataset, we replaced them with perceived-transcripts generated using Whisper (\emph{base.en})~\cite{radford2023robust}. Because GOPT only supports a maximum input length of 50 phonemes, utterances exceeding this limit were segmented into 50-phoneme chunks, and the final prediction was obtained by averaging scores across all chunks. For MultiPA, we also used the released checkpoint and code for evaluation. For Joint-APA, we followed the original training recipe, using a \emph{wav2Vec2-large-robust} encoder with a CTC loss weight of 1.0, a classification loss weight of 0.25, and classification-loss warmup beginning at epoch 50. Since Joint-APA frames assessment as an 11-class classification task over scores from 0 to 10, we computed the final prediction as the expectation over the softmax-normalized class probabilities. Performance was evaluated using the Pearson correlation coefficient (PCC). 

We selected Speechocean as the primary comparison dataset due to its widespread use in prior studies (e.g., GOPT, MultiPA, and Joint-APA) and its similar rating granularity, which also includes utterance-level accuracy, fluency, and prosody.

\section{Results}
\subsection{OpenEnded Test Set Statistics}

First, we evaluated inter-annotator agreement using Krippendorff’s $\alpha$. The resulting scores were 0.621, 0.732, and 0.637 for accuracy, fluency, and prosody, respectively. Fluency achieved the highest inter-annotator agreement, probably because perceptions of fluency are strongly influenced by salient cues such as unnaturally long pauses and filler words (e.g., “um”). These signals are more directly observable in the speech signal and thus less subjective to evaluate than accuracy or prosody. Accuracy reflects multiple factors, including grammar, pronunciation and vocabulary usage. While grammatical and phonetic errors can often be identified consistently, the intended meaning of a speaker’s response may remain ambiguous in open-response settings. As a result, annotators may differ in their judgments of whether a speaker’s vocabulary usage successfully conveys the message. For prosody, interpretations of a speaker’s rhythm, stress, and intonation may vary across annotators, as these factors are closely related and strongly influence one another. Annotators may also differ in the relative importance they assign to each prosodic factor when evaluating a given utterance. Nevertheless, the inter-annotator agreement indicates moderate consistency, suggesting that the annotations in the OpenEnded test set are reasonably reliable.

\begin{figure}[hptb!]
\centerline{\includegraphics[scale=0.86]{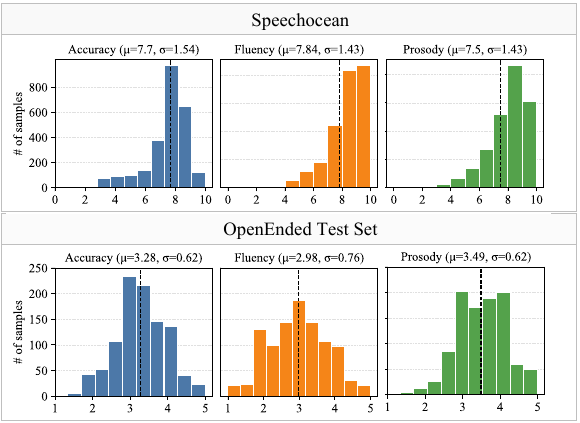}}
\caption{Score distributions of Speechocean and OpenEnded.}
\label{fig:distribution_comparison}
\end{figure}

The score distribution comparison in Fig.~\ref{fig:distribution_comparison} shows that Speechocean exhibits a noticeably left-skewed distribution, while the distributions in OpenEnded are closer to normal. In OpenEnded, fluency has the lowest mean score and the highest variance ($2.98 \pm 0.76$) compared with accuracy ($3.28 \pm 0.62$)  and prosody ($3.49 \pm 0.62$). In contrast, fluency scores in Speechocean are relatively higher and do not exhibit a particularly large standard deviation. This pattern may be attributed to the increased cognitive load in the open-response setting, where speakers must formulate content while simultaneously producing speech, thereby making fluent speech more challenging and amplifying differences in English proficiency. 

\begin{figure}[htbp!]
\centerline{\includegraphics[scale=0.75]{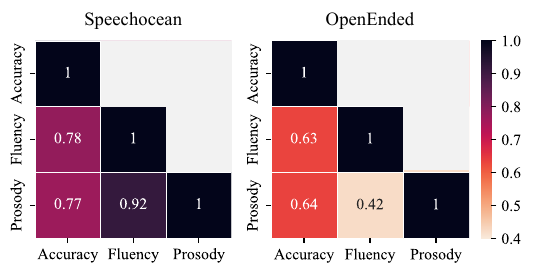}}
\caption{Inter-dimensional correlation of Speechocean and OpenEnded.}
\label{fig:confusion_comparision}
\end{figure}

Next, we examine inter-dimensional correlations within each dataset. As shown in Fig.~\ref{fig:confusion_comparision}, OpenEnded exhibits substantially lower correlations among assessment dimensions than Speechocean. This highlights a key difference in the scope of the underlying dimension definitions: although both datasets evaluate accuracy,  fluency, and prosody, OpenEnded defines these dimensions more broadly. In addition, the open-response setting offers greater potential for differentiating speaking abilities than read-aloud speech. As a result, the OpenEnded dimensions are more clearly separated while still exhibiting moderate correlations, reflecting overall speaking ability.

\subsection{Zero-Shot and Cross-Dataset Performance}
We report performance on the OpenEnded test set without using OpenEnded training data, including zero-shot evaluation of ALMs and cross-dataset evaluation of ASA models trained on Speechocean (Table~\ref{tab:zero_shot_baseline}). For ALMs, we observe that Gemini-2.0 shows no clear performance advantage between single-task and multi-task settings. We further find that Gemini-2.5 does not exhibit clear performance improvements over Gemini-2.0 on OpenEnded, despite its stronger general capabilities. For ASA models, the proposed VoxPA outperforms previous models in fluency and prosody, despite slightly lower accuracy than Joint-APA. Lastly, we use ALM performance as a reference for evaluating ASA models in an out-of-domain setting. While ASA models outperform Gemini in accuracy and prosody, they under-perform in fluency. This may reflect a limitation of training on a read-aloud dataset such as Speechocean and applying the model to an open-response setting, where fluency characteristics differ. 

We also evaluated open-source ALMs, including Qwen~\cite{chu2024qwen2} and Flamingo~\cite{ghosh2026audio}, using their \emph{qwen2-audio-7B-instruct} and \emph{audio-flamingo-3-hf} models, respectively. However, both substantially underperformed Gemini, showing limited ability in the zero-shot setting. To enhance dataset utility, we selected Gemini (\emph{i.e., gemini-2.0-AFP}) to generate pseudo-labels for the remaining data.

\begin{table}[t]
\caption{Zero-Shot and Cross-Dataset Performance}
\begin{center}
\begin{tabular}{|cccc|}
\hline
\multicolumn{1}{|c|}{\textbf{Model}} & \multicolumn{1}{c|}{\cellcolor[HTML]{F1F1F1}{\textbf{Accuracy}}} & \multicolumn{1}{c|}{\cellcolor[HTML]{F1F1F1}{\textbf{Fluency}}} & {\cellcolor[HTML]{F1F1F1}\textbf{Prosody}} \\ \hline \hline
\multicolumn{4}{|c|}{\cellcolor[HTML]{F1F1F1}{ALM} }                                                                              \\ \hline
\multicolumn{1}{|c|}{gemini-2.0-A}   & \multicolumn{1}{c|}{0.414}     & \multicolumn{1}{c|}{-}      & -      \\ \hline
\multicolumn{1}{|c|}{gemini-2.0-F}   & \multicolumn{1}{c|}{\textbf{-}} & \multicolumn{1}{c|}{0.437} & -      \\ \hline
\multicolumn{1}{|c|}{gemini-2.0-P}   & \multicolumn{1}{c|}{\textbf{-}} & \multicolumn{1}{c|}{-}      & 0.346 \\ \hline
\multicolumn{1}{|c|}{gemini-2.0-AFP} & \multicolumn{1}{c|}{0.403}      & \multicolumn{1}{c|}{0.451}  & 0.339  \\ \hline
\multicolumn{1}{|c|}{gemini-2.5-AFP} & \multicolumn{1}{c|}{0.348}      & \multicolumn{1}{c|}{0.479}  & 0.324  \\ \hline
\multicolumn{4}{|c|}{\cellcolor[HTML]{F1F1F1}ASA Model (training data: Speechocean)}                          \\ \hline
\multicolumn{1}{|c|}{MultiPA}        & \multicolumn{1}{c|}{0.393}      & \multicolumn{1}{c|}{0.405}  & 0.400  \\ \hline
\multicolumn{1}{|c|}{GOPT}           & \multicolumn{1}{c|}{0.217}      & \multicolumn{1}{c|}{0.300}  & 0.213  \\ \hline
\multicolumn{1}{|c|}{Joint-APA}      & \multicolumn{1}{c|}{\textbf{0.440} }     & \multicolumn{1}{c|}{0.418}  & 0.423  \\ \hline \hline
\multicolumn{1}{|c|}{VoxPA}          & \multicolumn{1}{c|}{0.421}     & \multicolumn{1}{c|}{\textbf{0.451}} & \textbf{0.456} \\ \hline
\end{tabular}
\label{tab:zero_shot_baseline}
\end{center}
\end{table}

\subsection{ASA Model Training with ALM Pseudo-labels}

Table~\ref{tab:pseudo_training} presents the performance improvements obtained by training ASA models with ALM-generated pseudo-labels, i.e., the OpenEnded training and development set. Note that the reported performance of \emph{gemini-2.0-AFP}, the ALM used to generate the pseudo-labels, reflects the correlation between its scores and human annotations on the test set, indicating the quality of ALM pseudo-labels. We evaluate Joint-APA trained from scratch on OpenEnded (denoted as Joint-APA), Joint-APA initialized from Speechocean and further fine-tuned on OpenEnded (denoted as Joint-APA\textsubscript{SO}), and the proposed VoxPA under the same experimental setting. The experimental results show that all models achieve performance gains when trained on OpenEnded pseudo-labeled data, demonstrating that ALM-based pseudo-labeling is a potential alternative when manual annotation is costly or limited. Notably, the proposed VoxPA consistently outperforms Joint-APA, with the most pronounced gains observed in fluency, establishing it as a stronger baseline on the OpenEnded dataset. Lastly, both models show that pretraining on Speechocean before training on OpenEnded performs better than training directly on OpenEnded. However, due to the high internal correlation within Speechocean, pretraining on it may increase inter-dimension correlation, leading different prediction dimensions to capture similar rather than distinct aspects of speaking proficiency.

\begin{table}[hpbt!]
\caption{Performance of ASA Models Trained with ALM Pseudo-Labels}
\begin{center}
\begin{tabular}{|cccc|}
\hline 
\multicolumn{1}{|c|}{Model} &
  \multicolumn{1}{c|}{\cellcolor[HTML]{F1F1F1}{\textbf{Accuracy}}} &
  \multicolumn{1}{c|}{\cellcolor[HTML]{F1F1F1}{\textbf{Fluency}}} &{\cellcolor[HTML]{F1F1F1}
  \textbf{Prosody} }\\ \hline \hline
\multicolumn{1}{|c|}{gemini-2.0-AFP} & \multicolumn{1}{c|}{0.403}  & \multicolumn{1}{c|}{0.451}  & 0.339  \\ \hline
\multicolumn{1}{|c|}{\begin{tabular}[c]{@{}c@{}}Joint-APA\\ (Speechocean)\end{tabular}} &
  \multicolumn{1}{c|}{0.440} &
  \multicolumn{1}{c|}{0.418} &
  0.423 \\ \hline \hline
\multicolumn{4}{|c|}{\cellcolor[HTML]{F1F1F1}{OpenEnded Pseudo-label Training} }                              \\ \hline \hline
\multicolumn{1}{|c|}{Joint-APA}      & \multicolumn{1}{c|}{0.480}  & \multicolumn{1}{c|}{0.508}  & 0.441  \\ \hline
\multicolumn{1}{|c|}{Joint-APA$_{SO}$}      & \multicolumn{1}{c|}{0.484}  & \multicolumn{1}{c|}{0.544}  & 0.484  \\ \hline
\multicolumn{1}{|c|}{VoxPA}          & \multicolumn{1}{c|}{0.506} & \multicolumn{1}{c|}{0.699}  & 0.464 \\ \hline
\multicolumn{1}{|c|}{VoxPA$_{SO}$}          & \multicolumn{1}{c|}{0.542} & \multicolumn{1}{c|}{0.703} & 0.529 \\ \hline
\end{tabular}
\label{tab:pseudo_training}
\end{center}
\end{table}

\subsection{Analysis of ALM for ASA}\label{sec:exp_alm_asa}

We analyze the ALM’s scores and representative reasoning on the OpenEnded training and development sets to evaluate whether its generated descriptions align with the assigned score levels and faithfully reflect the targeted assessment dimensions (Table~\ref{tab:description}). The representative reasoning is obtained by first computing sentence embeddings for all reasoning descriptions and then selecting the description closest to the centroid embedding of each group. First, although the ALM prompt does not provide specific definitions for each dimension, its generated descriptions are consistent with the assigned scores and reflect the intended characteristics: accuracy focuses on correctness and intelligibility, fluency on pauses and hesitation, and prosody on intonation and rhythm. The Measure of Textual Lexical Diversity (MTLD)\footnote{\url{https://pypi.org/project/lexical-diversity/}} for the reasoning of accuracy, fluency, and prosody are 83.96, 67.40, and 72.95, respectively. The higher lexical diversity in accuracy-related reasoning can be attributed to the fact that accuracy assessments often refer to specific words within an utterance, which vary across different responses. In contrast, fluency and prosody are more closely related to the overall speaking impression, leading to more general and repetitive patterns across utterances. 

\begin{table}[t]
\caption{Representative ALM Reasoning Across Score Levels}
\setlength{\tabcolsep}{3pt}
\begin{center}
\renewcommand{\arraystretch}{1.1}
\scalebox{0.9}{
\begin{tabularx}{1.05\columnwidth}{|l|>{\raggedright\arraybackslash}X|}
\hline 
\multicolumn{2}{|c|}{\cellcolor[HTML]{F1F1F1}{Accuracy}} \\ \hline
 \textbf{1} &  The input consists of sounds that are not speech. There are no recognizable words or phonemes.  \\ \hline
\textbf{2} &  The pronunciation contains noticeable errors.  \\ \hline
\textbf{3} &  The pronunciation is generally understandable, but there are a few errors. Some sounds are unclear, making comprehension slightly challenging.  \\ \hline
\textbf{4} &  The pronunciation is generally clear and understandable, with only minor errors. The speaker's articulation of individual words is mostly accurate.\\ \hline
\textbf{5} &  The pronunciation is completely accurate and natural, with no discernible errors. It mirrors a native speaker's pronunciation. \\ \hline \hline

\multicolumn{2}{|c|}{\cellcolor[HTML]{F1F1F1}{Fluency}} \\  \hline

\textbf{1} &  The speech is too fragmented and unclear due to the overwhelming background noise.  \\ \hline
\textbf{2} &  The speech is hesitant with frequent pauses and fillers (**um**). This significantly disrupts the flow of speech. \\ \hline
\textbf{3} &  The speech contains pauses and hesitations, which impact the fluency. \\ \hline
\textbf{4} &  The speech is mostly fluent, with only occasional hesitations. \\ \hline
\textbf{5} &  The speech flows smoothly and effortlessly, with appropriate pacing and natural pauses. \\ \hline \hline 

\multicolumn{2}{|c|}{\cellcolor[HTML]{F1F1F1}{Prosody}} \\  \hline
\textbf{1} &  Due to the absence of discernible speech, prosody (intonation, stress, rhythm) cannot be evaluated. \\ \hline
\textbf{2} &  The intonation is somewhat flat and lacks natural variation. 
\\ \hline
\textbf{3} &  The intonation and stress patterns are somewhat monotonous. \\ \hline
\textbf{4} &  Intonation and stress are generally accurate and varied, contributing to natural-sounding speech.\\ \hline
\textbf{5} &  The speaker uses intonation, stress, and rhythm effectively to convey meaning and engage the listener. \\ \hline 
\hline 
\end{tabularx}
}
\label{tab:description}
\end{center}
\end{table}

\begin{figure}[htbp!]
\centerline{\includegraphics[scale=0.85]{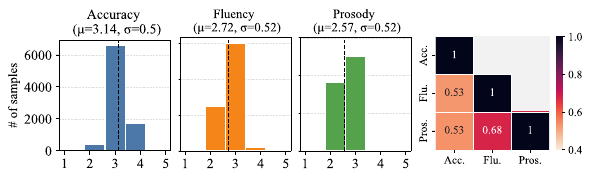}}
\caption{Score distribution and inter-dimension correlation.}
\label{fig:train_dev_distribution}
\end{figure}

Fig.~\ref{fig:train_dev_distribution} presents the score distributions. The ALM exhibits moderate inter-dimension correlations, similar to those observed in human annotations. However, it shows a stronger tendency to assign mid-range scores (i.e., score 3), while extreme scores (1 and 5) are rarely assigned. According to the model’s reasoning, score 1 is typically assigned when no speech is detected, which may occur in real-world data collection when users do not speak after initiating recording. This pattern may also reflect the dataset, in which most students have sufficient English proficiency to attempt open responses rather than being complete beginners. In contrast, the ALM rarely assigns a score of 5, as speakers are non-native learners still developing their English proficiency and thus unlikely to exhibit native-like performance. Compared with human annotators (Fig.~\ref{fig:distribution_comparison}), humans assign a higher proportion of extreme scores, consistent with prior studies~\cite{cooper2023investigating} reporting that human raters tend to utilize the full scoring range. That is, human raters may form expectations over the dataset and calibrate their scoring accordingly, resulting in full-range usage, whereas the ALM evaluates each sample independently.

Overall, the analysis of the ALM's reasoning suggests that it captures meaningful distinctions across assessment dimensions and produces generally coherent evaluations. Therefore, ALMs also have the potential to support ASA development with descriptive feedback in addition to pseudo-label scores. However, manual screening of such reasoning descriptions remains a limitation of this study, along with the lack of high-quality reasoning annotations, models capable of generating reliable descriptive feedback, and established evaluation protocols for assessing generated descriptions.

\section{Conclusion}
We introduce OpenEnded, an ASA dataset that uniquely provides utterance-level scoring for open-response speech. Speaking proficiency is evaluated along three dimensions by human annotators to construct a test set and ALM pseudo-labels are used to establish the development and training set. We further establish benchmark results on the annotated data, in which the proposed VoxPA trained on the ALM pseudo-labeled data achieved the best performance. Lastly, we analyze the descriptive feedback generated by the ALM and observe that accuracy-related feedback is more dependent on utterance content and typically provides more specific details, whereas fluency and prosody feedback tend to be more generic. In summary, our study introduces a new manually annotated ASA benchmark, demonstrates the effectiveness of ALM-generated pseudo-labels, and highlights the potential to leverage ALMs for constructing speaking proficiency assessment datasets with descriptive feedback.

\bibliographystyle{IEEEtran}
\bibliography{refs}

\end{document}